\documentclass{article}
\usepackage{spconf,amsmath,graphicx,hyperref}

\usepackage{fix-cm}
\usepackage{amsmath}   
\usepackage{booktabs}  
\usepackage{graphicx}  
\usepackage{multirow}
\usepackage{xcolor}
\usepackage[fontsize=9pt]{fontsize}

\usepackage[numbers,sort&compress]{natbib}
\let\OLDthebibliography\thebibliography
\renewcommand\thebibliography[1]{
  \OLDthebibliography{#1}
  \setlength{\parskip}{4pt}    
  \setlength{\itemsep}{-0.5ex} 
}

\title{E\lowercase{mpowering} M\lowercase{ultimodal} R\lowercase{espiratory} S\lowercase{ound} C\lowercase{lassification} \lowercase{with} \\ C\lowercase{ounterfactual} A\lowercase{dversarial} D\lowercase{ebiasing} \lowercase{for} O\lowercase{ut}-\lowercase{of}-D\lowercase{istribution} R\lowercase{obustness}}

\name{Heejoon Koo$^{1, 2}$$\thanks{\hspace{-1em}$^{\dagger}$ corresponding author. This work was supported by the National Research Foundation of Korea(NRF) grant funded by the Korea government(MSIT) (Grant no. RS-2025-16066662).}$, Miika Toikkanen$^{2}$, Yoon Tae Kim$^{2, 3}$, Soo Yong Kim$^{2}$, June-Woo Kim$^{\dagger2,4}$}
\address{$^{1}$University College London, United Kingdom, $^{2}$RSC LAB, Republic of Korea  \\ $^{3}$KAIST, Republic of Korea, $^{4}$AI-ACE InnoCORE, GIST, Republic of Korea}

\begin{document}

\maketitle

\begin{abstract}

Multimodal respiratory sound classification offers promise for early pulmonary disease detection by integrating bioacoustic signals with patient metadata. Nevertheless, current approaches remain vulnerable to spurious correlations from attributes such as age, sex, or acquisition device, which hinder their generalization, especially under distribution shifts across clinical sites. To this end, we propose a counterfactual adversarial debiasing framework. First, we employ a causal graph-based counterfactual debiasing methodology to suppress non-causal dependencies from patient metadata. Second, we introduce adversarial debiasing to learn metadata-insensitive representations and reduce metadata-specific biases. Third, we design counterfactual metadata augmentation to mitigate spurious correlations further and strengthen metadata-invariant representations. By doing so, our method consistently outperforms strong baselines in evaluations under both in-distribution and distribution shifts. Code is available at \textcolor{cyan}{\href{https://github.com/RSC-Toolkit/BTS-CARD}{https://github.com/RSC-Toolkit/BTS-CARD}}.

\end{abstract}


\section{Introduction}
\label{sec:introduction}

Respiratory sound classification (RSC) provides a non-invasive and cost-effective method for the early detection of pulmonary diseases, offering a practical alternative to manual auscultation \cite{sarkar2015auscultation, bae2023patch, kim2024bts}. Clinicians typically base their diagnostic decisions on heterogeneous information sources, including patient metadata \cite{kim2024bts, koo2024next, krones2025review}. Motivated by this clinical reasoning process, recent advances in RSC have leveraged multimodal learning to integrate respiratory audio with auxiliary metadata (age, sex, stethoscope, etc.) \cite{kim2024bts}. 

While such multimodal systems have substantially improved RSC performance, several critical challenges persist. First, the use of patient metadata can introduce spurious correlations, leading to biased predictions \cite{varoquaux2022machine}. For example, pediatric cases are frequently over-represented among healthy samples, potentially inducing age-related confounding when learning representations. Second, models often fail to generalize across clinical environments that differ in stethoscopes (recording devices) or measurement protocols, limiting their real-world deployment \cite{kim2024stethoscope, lasko2024probabilistic, koo2024next, kim2025adaptive}. These issues may undermine generalization, posing significant obstacles to real-world adoption.

Causal inference provides a principled framework to address such limitations by estimating the effect of specific variables while controlling for confounders \cite{pearl2009causal, scholkopf2021toward}. This perspective is especially valuable in multimodal learning, where multimodal integration can make it more challenging to identify the true causal mechanisms, while also amplifying modality-driven biases \cite{cadene2019rubi, niu2021counterfactual}. In RSC, metadata often exhibit skewed distributions across diagnostic categories, leading models to rely on superficial correlations instead of clinically meaningful features \cite{varoquaux2022machine, yang2022machine}. Counterfactual debiasing mitigates this by disentangling causal signals from spurious ones, thereby enhancing in-distribution (IND) performance and out-of-distribution (OOD) robustness \cite{scholkopf2021toward, kaddour2022causal}.

Generalization is critical for clinical AI systems to ensure reliable performance across heterogeneous hospitals and patient populations \cite{yang2022machine, koo2023survey, koo2023comprehensive}. Beyond causal methods, robustness under OOD conditions requires learning representations that remain stable across variations in demographics, acquisition devices, and clinical protocols \cite{vaidya2024demographic}. Adversarial debiasing, originally introduced for algorithmic fairness, can be adapted to suppress hospital-specific attributes by jointly optimizing the main prediction task and adversarial domain classifiers \cite{zhang2018mitigating, li2018deep, yang2023adversarial}. Such training encourages metadata-agnostic representations and improves cross-site generalization \cite{li2018deep}. However, existing research on RSC has overlooked the integration of counterfactual reasoning with adversarial debiasing to improve robustness under distribution shifts.

\begin{figure*}[!ht]
    \centering
    \includegraphics[width=0.95\textwidth]{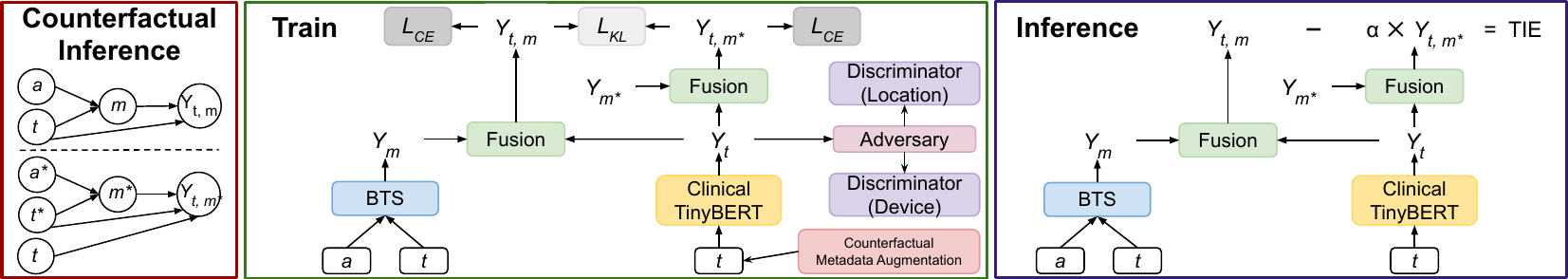}
    \caption{An Overview of Our Counterfactual Adversarial Debiasing Framework for Multimodal RSC, BTS-CARD.}
    \label{fig:overview_framework}
    \vspace{-5mm}
\end{figure*}

Therefore, we propose BTS-CARD, a novel counterfactual adversarial debiasing framework for multimodal RSC (see Figure \ref{fig:overview_framework}). First, we introduce a counterfactual debiasing that suppresses non-causal associations, yielding more reliable predictions in both IND (ICBHI \cite{rocha2017alpha}) and OOD (SPRSound \cite{zhang2022sprsound}) settings. Second, we enhance generalization via adversarial debiasing, which promotes metadata-invariant representations and prevents encoding acquisition device and recording location. Third, we devise counterfactual metadata augmentation to further mitigate spurious correlations from patient metadata and strengthen metadata-agnostic representations by neutralizing sensitive attributes. Extensive experiments demonstrate that our method consistently outperforms strong baselines, validating its robustness and transferability.

\section{Methodologies}
\label{sec:methodologies}

\subsection{Preliminaries on Causal Inference}
\label{sec:preliminaries}

\begin{figure}[ht]
    \centering
    \includegraphics[width=0.9\columnwidth]{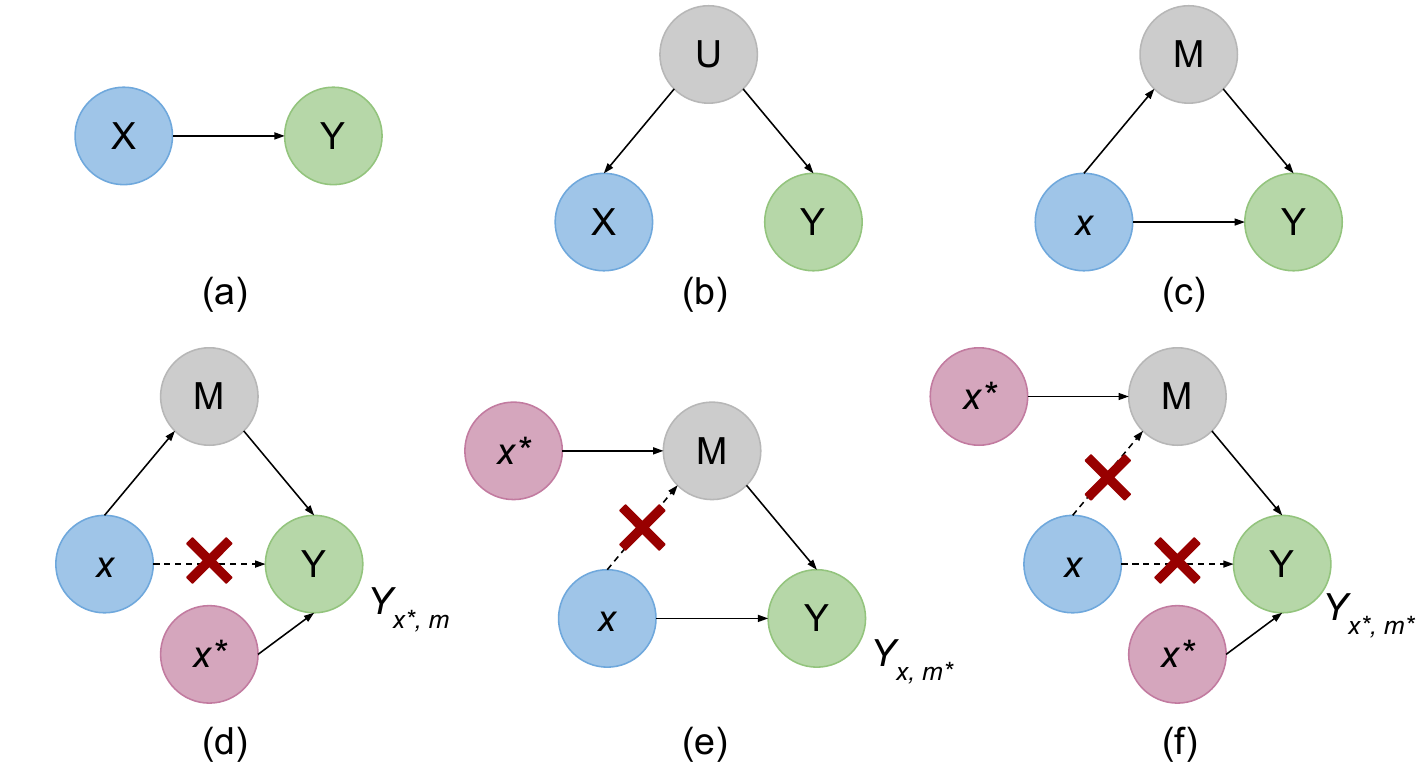}
    \caption{Causal Graph.}
    \label{fig:causal_graph}
    \vspace{-3mm}
\end{figure}

\noindent\textbf{Causal Graph.}  
A causal graph represents dependencies between variables as a directed acyclic graph $G = \{N, E\}$, where $N$ is the set of variables (nodes) and $E$ is the set of directed edges denoting causal relations \cite{pearl2009causal, kaddour2022causal}. For example, in Figure~\ref{fig:causal_graph}(a), variable $X$ directly influences $Y$. Confounding occurs when a hidden variable $U$ influences both $X$ and $Y$ (Figure~\ref{fig:causal_graph}(b)), making it impossible to attribute their observed correlation to a direct causal effect.

Counterfactual reasoning evaluates outcomes under hypothetical interventions that differ from the observed world \cite{pearl2009causal}. In the factual world, the mediator $M$ naturally depends on $X$, leading to outcomes $Y_{x,m}$ (Figure~\ref{fig:causal_graph}(c)). Counterfactual settings alter $X$ or $M$ while fixing others. For instance, the following describes the outcome when $X$ is set to $x^\ast$ but $M$ is fixed at the value induced by $X=x$ (Figure~\ref{fig:causal_graph}(d)). 
\[
Y_{x^\ast, m} = Y(X = x^\ast, m = M(X = x)).
\]  

\noindent\textbf{Causal Intervention.}  
To formally identify causal effects, Pearl’s $do$-calculus \cite{pearl2009causal} prescribes rules for reasoning about interventions. An intervention $do(X=x)$ severs all incoming edges into $X$, ensuring its value is externally fixed rather than caused by its parents. For instance, in Figure~\ref{fig:causal_graph}(b), computing $P(X \mid do(Y))$ removes the confounding effect of $U \to X$, thereby isolating the true causal effect of $X$ on $Y$. This principle forms the foundation of our debiasing approach, which suppresses spurious dependencies while preserving genuine causal mechanisms.

\noindent\textbf{Causal Effects.}  
Causal effects measure differences in potential outcomes under different treatments \cite{pearl2009causal}. The Total Effect (TE) of $X$ on $Y$ compares the outcome under $X=x$ (Figure~\ref{fig:causal_graph}(c)) to that under $X=x^\ast$ (Figure~\ref{fig:causal_graph}(f)):  
\begin{equation}
\text{TE} = Y_{x, m} - Y_{x^\ast, m^\ast}.
\end{equation}  
TE can be decomposed into a Natural Direct Effect (NDE) and a Total Indirect Effect (TIE). The NDE isolates the direct path from $X$ to $Y$ by fixing the mediator $M$ at $m\ast$ (Figure~\ref{fig:causal_graph}(e)):  
\begin{equation}
\text{NDE} = Y_{x, m^\ast} - Y_{x^\ast, m^\ast}.
\end{equation}  
The TIE captures the mediated contribution via $M$ (Figure~\ref{fig:causal_graph}(d)):  
\begin{equation}
\text{TIE} = \text{TE} - \text{NDE} = Y_{x, m} - Y_{x, m^\ast}.
\end{equation}  
We adopt TIE for the unbiased inference \cite{pearl2009causal, niu2021counterfactual, xu2023counterfactual, vosoughi2024cross}. Such decomposition clarifies the roles of direct and mediated pathways and offers a principled means to mitigate bias. Quantifying mediated effects is especially vital in domains like healthcare, where hidden pathways may encode spurious factors.

\subsection{Task Formulation}
\label{sec:task_formulation}

\noindent\textbf{Traditional Multimodal RSC.} A joint representation $m$ is constructed from audio $a$ and patient metadata $t$, which is then mapped to lung sound predictions $Y_{t,m}$ (the top half of the Counterfactual Inference in Figure~\ref{fig:overview_framework}). Without modeling the underlying causal structure, such approaches conflate direct and indirect effects of metadata and often exploit shortcut paths such as $t \rightarrow Y_{t,m}$, resulting in biased and non-generalizable predictions.

\noindent\textbf{Debiased Multimodal RSC.} From a causal perspective, we decompose the influence of metadata into two pathways: the spurious direct path $t \rightarrow Y_{t,m}$ and the informative indirect path $(a, t) \rightarrow m \rightarrow Y_{t,m}$. Our framework employs counterfactual debiasing, illustrated by contrasting the top (factual) and bottom (counterfactual) graphs in the Counterfactual Inference of Figure~\ref{fig:overview_framework}, to disentangle causal effects from confounding. Combined with adversarial debiasing and counterfactual metadata augmentation, our approach further learns representations resilient to metadata variance, thereby improving accuracy and robustness against distribution shifts.

\subsection{Counterfactual Adversarial Debiasing Framework}
\label{sec:rsc_framework}

In this section, we present the counterfactual adversarial framework for multimodal RSC that builds upon the BTS model \cite{kim2024bts}. 

\noindent\textbf{Counterfactual Debiasing.} To mitigate spurious influences of metadata and achieve unbiased inference, we propose a counterfactual debiasing framework. We begin by estimating the TE of BTS by contrasting the factual and counterfactual worlds:
\begin{equation}
\text{TE} = Y_{t,m} - Y_{t^{*},m^{*}},
\end{equation}
where $Y_{t,m}$ denotes the prediction under text derived by metadata prompt $t$ \cite{kim2024bts} and multimodal representation $m$, while $Y_{t^{*},m^{*}}$ corresponds to their counterfactual counterparts. Here, multimodal counterfactual features are instantiated using a dummy vector $m^{*}$, fixed to a constant value of 1 based upon empirical validation. We use RUBi \cite{cadene2019rubi} for fusion, given its established effectiveness in causal modeling. Next, we estimate the NDE of metadata text using Clinical TinyBERT \cite{rohanian2024lightweight}, which is not only efficient, but also easily adaptable to new tasks via fine-tuning. The NDE isolates the spurious shortcut $T \to Y$, and the TIE is then obtained by subtracting NDE from TE:
\begin{equation}
\begin{alignedat}{2}
    & \text{NDE} &&= Y_{t,m^{*}} - Y_{t^{*},m^{*}}, \\
    & \text{TIE} &&= \text{TE} - \text{NDE} = Y_{t,m} - Y_{t,m^{*}}.
\end{alignedat}
\end{equation}

At inference, we adopt TIE as the debiased prediction, ensuring that model outputs leverage the informative contributions of metadata while suppressing spurious correlations.

\noindent\textbf{Adversarial Debiasing.} 
To promote metadata-agnostic representations for robustness across different clinical environments, we apply adversarial debiasing \cite{zhang2018mitigating} to the NDE. To mitigate deployment bias, both recording devices and locations are targeted as they vary across clinical sites.

Debiasing is restricted to the natural direct effect (NDE), with logits $Y_{t}$ fed to an adversary $U$ to yield representation $z$. A Gradient Reversal Layer (GRL, $ \mathcal{R} $) between $z$ and $D_a$ reverses gradients to promote metadata-insensitive features, with both tasks optimized via CE. Thus, with $D_{a}$, the adversarial discriminator for the metadata attribute $a \in \{\text{location}, \text{device}\}$ and $l_{a}$, its corresponding ground-truth label, the objective is as follows:
\begin{equation}
\begin{alignedat}{3}
L_{\text{adv}} &= L_{\text{CE}_\text{NDE}} 
+ \sum_{a \in \{\text{location}, \text{device}\}} \lambda_{a} L_{a}, \\
\text{where} \: L_{\text{CE}_{\text{NDE}}} &= \text{CE}(Y_{t}, y) \: \text{and} \: L_{a} = \text{CE}(D_{a}(\mathcal{R}(z)), l_{a}).
\end{alignedat}
\end{equation}

\noindent\textbf{Counterfactual Data Augmentation.} To encourage invariance in metadata representations, we replace sensitive metadata with neutral placeholders (e.g., `This patient is an adult patient.' → `This patient’s age is unknown.') in the NDE model (Clinical TinyBERT \cite{rohanian2024lightweight}), instead of simply erasing or converting to UNK token \cite{zhong2020random, koo2025overcoming}.

From a causal perspective, this intervention suppresses the spurious shortcut $T \to Y$ while leaving the TE model intact to preserve both direct and indirect effects \cite{scholkopf2021toward}. This methodology enforces insensitivity to non-causal attributes without discarding meaningful pathways. It parallels the logic of $do$-calculus \cite{pearl2009causal}, where interventions isolate causal effects by severing spurious dependencies. Replacing sensitive metadata in this manner thus mimics a $do$-operation that removes the $T \to Y$ link while preserving potential mediating paths to better reflect the hypothesized causal structure.

\noindent\textbf{Training and Inference.}
For training, we optimize both factual ($Y_{t,m}$) and counterfactual ($Y_{t,m^{*}}$) predictions using two CE losses:
\begin{equation}
L_{\text{CE}} = \text{CE}(Y_{t,m}, y) + \text{CE}(Y_{t,m^{*}}, y).
\end{equation}
To stabilize counterfactual estimation, we add a KL divergence term that aligns $Y_{t,m^{*}}$ with $Y_{t,m}$:
\begin{equation}
L_{\text{KL}} = \text{KL}(Y_{t,m^{*}} \,\|\, Y_{t,m}).
\end{equation}
Thus, with $\lambda_{\text{CE}}$ and $\lambda_{\text{KL}}$ controlling the weights of CE and KL regularization, the overall training loss is:
\begin{equation}
L = \lambda_{\text{CE}} L_{\text{CE}} + \lambda_{\text{KL}} L_{\text{KL}} + L_{\text{adv}}.
\end{equation}

\noindent At inference, we debias by subtracting the estimated NDE from the TE, retaining only the TIE. The final prediction is:
\begin{equation}
\hat{Y} = Y_{t,m} - \alpha \cdot Y_{t,m^{*}} = \text{TIE},
\end{equation}
where $\alpha$ controls the degree of direct-effect removal, thereby effectively suppressing spurious metadata shortcuts and promoting more generalizable predictions.

\section{Experiments}
\label{sec:experiments}

\subsection{Experimental Setup}
\label{sec:experimental_setup}

\noindent\textbf{Dataset.} We use two datasets: 1) ICBHI Respiratory Sound Database \cite{rocha2017alpha}, for IND set, and 2) SPRSound \cite{zhang2022sprsound}, the Shanghai Jiao Tong University (SJTU) Pediatric Respiratory Sound Database, for OOD test set. First, ICBHI is annotated into four classes. Following the BTS \cite{kim2024bts}, we binarize age into pediatric and adult groups and retain all other metadata as per the original ICBHI dataset specification. Second, SPRSound originally contains seven classes; to align with ICBHI, we merge crackle-related labels into crackle and combine Stridor and Rhonchi into wheeze. We also use the inter-patient-level validation set only for OOD test set. Comprehensive details are summarized in Table \ref{tab:datasets}.

\noindent\textbf{Training Details.} Following BTS \cite{kim2024bts}, we extract respiratory cycles, standardize them to 8 s, and resample the audio to 48 kHz. Text descriptions are capped at 64 tokens, which are sufficient to encode all text metadata without truncation. In line with BTS, we use all metadata and avoid relying solely on common attributes (e.g., sex) across datasets, ensuring the best IND performance.

For counterfactual debiasing, we set $\lambda_{\text{CE}}=1.0$ and $\lambda_{\text{KL}}=1.0$, respectively. We vary the value of $\alpha$ from 0 to 1 in increments of 0.1 to analyze the effect of $\alpha$ on overall performance. During training, counterfactual metadata augmentation is employed by independently replacing each metadata attribute with a neutral placeholder with a probability $ p $ of 0.25. For adversarial debiasing, the adversary is applied with the coefficient of 1.0, while the location and device discriminator losses are weighted by $\lambda_{\text{location}} = 0.01$ and $\lambda_{\text{device}} = 0.1$, respectively. We fine-tune using the AdamW optimizer \cite{loshchilov2017decoupled}, with a learning rate of $5\times10^{-5}$ for all parameters. It is further decayed via cosine scheduling over 30 epochs with a batch size of 8. 

\noindent\textbf{Evaluation Protocol.} We use Specificity ($S_p$), Sensitivity ($S_e$), and their arithmetic mean (ICBHI Score) \cite{rocha2017alpha}. $S_p$ is the proportion of normal cases, and $S_e$ is that of abnormal cases correctly classified. Results are reported as the mean and variance over five runs. We use PyTorch \cite{paszke2019pytorch} and a single NVIDIA RTX 3090 for all experiments.

\begin{table}[!t]
  \centering
  \caption{Details of the ICBHI and SPRSound datasets. L/R denotes left or right. `Both' label is co-occurrence of crackle and wheeze.}
  \scriptsize
  \renewcommand{\arraystretch}{1.2}
  \setlength{\tabcolsep}{5pt}
  \begin{tabular}{@{} c c c c @{}}
    \toprule
    \midrule
    Dataset & Criteria & Type & Characteristics \\
    \hline
    \midrule
    \multirow{10}{*}{ICBHI}
      & \multirow{5}{*}{Metadata} & Age         & Adult, Pediatric \\
      &                            & Sex         & Male, Female \\
      &                            & Location    & Trachea, L/R Anterior, L/R Posterior, L/R Lateral \\
      &                            & Stethoscope & Meditron, LittC2SE, Litt3200, AKGC417L \\
      &                            & Others      & BMI (Adult only), Weight/Height (Pediatric only) \\
      \cline{2-4}
      & \multirow{5}{*}{Class Dist.}
        & Label / Ratio &
          \multirow{5}{*}{%
            \begin{tabular}{@{}c c c@{}}
              Train & Valid & Overall \\
              \hline
              2063 (49.81\%) & 1579 (57.29\%) & 3642 (52.80\%) \\
              1215 (29.33\%) & 649 (23.55\%)  & 1864 (27.02\%) \\
              501 (12.10\%)  & 385 (13.97\%)  & 886 (12.84\%)  \\
              363 (8.76\%)   & 143 (5.19\%)   & 506 (7.34\%)   \\
            \end{tabular}
          } \\
      & & Normal & \\
      & & Crackle & \\
      & & Wheeze  & \\
      & & Both    & \\
    \midrule
    \multirow{9}{*}{SPRSound}
      & \multirow{4}{*}{Metadata} & Age         & Pediatric \\
      &                            & Sex         & Male, Female \\
      &                            & Location    & L/R Anterior, L/R Posterior \\
      &                            & Stethoscope & Yunting model II \\
      \cline{2-4}
      & \multirow{5}{*}{Class Dist.}
        & Label / Ratio &
          \multirow{5}{*}{%
            \begin{tabular}{@{}c c c@{}}
              Train & Valid & Overall \\
              \hline
              5159 (78.14\%) & 1040 (72.78\%) & 6199 (77.19\%) \\
              961 (14.56\%)  & 83 (5.81\%)    & 1044 (13.00\%) \\
              452 (6.85\%)   & 305 (21.34\%)  & 757 (9.43\%)   \\
              30 (0.45\%)    & 1 (0.70\%)     & 31 (0.39\%)    \\
            \end{tabular}
          } \\
      & & Normal & \\
      & & Crackle & \\
      & & Wheeze  & \\
      & & Both    & \\
    \bottomrule
  \end{tabular}
  \label{tab:datasets}
  \vspace{-5mm}   
\end{table}

\subsection{Experimental Results}
\label{sec:experimental_results}

\subsubsection{Main Results}
\label{sec:main_results}

\begin{table*}[!t]
    \centering
    \caption{Main Results on IND (ICBHI) and OOD (SPRSound) settings. Best results are in boldface and the second-best results are underlined.}
    \renewcommand{\arraystretch}{1}
    \addtolength{\tabcolsep}{6pt}
    \resizebox{\linewidth}{!}{
    \begin{tabular}{ccccccccc}
    \toprule
    \midrule
    \multirow{2}{*}{Criteria} & \multirow{2}{*}{Method} & \multirow{2}{*}{Venue} 
    & \multicolumn{3}{c}{IND (In-Distribution)} 
    & \multicolumn{3}{c}{OOD (Out-of-Distribution)} \\
    \cmidrule(lr){4-6}\cmidrule(lr){7-9}
    & & & $S_p$\,(\%) & $S_e$\,(\%) & Score\,(\%) & $S_p$\,(\%) & $S_e$\,(\%) & Score\,(\%) \\
    \hline \midrule

    \multirow{6}{*}{Unimodal} 
    & Moummad \textit{et al.} \cite{moummad2023pretraining} & \textit{WASPAA`23} & 70.09 & 40.39 & 55.24 & -- & -- & -- \\
    & Moummad \textit{et al.} \cite{moummad2023pretraining} \,(SCL) & \textit{WASPAA`23} & 75.95 & 39.15 & 57.55 & -- & -- & -- \\    
    & Bae \textit{et al.} \cite{bae2023patch} \, (Fine-tuning) & \textit{INTERSPEECH`23} & 77.14 & 41.97 & 59.55 & 69.62 & 32.65 & 51.13 \\
    & Bae \textit{et al.} \cite{bae2023patch} \, (Patch-Mix CL) & \textit{INTERSPEECH`23} & 81.66 & 43.07 & 62.37 & 62.69 & \underline{39.33} & 51.01 \\
    & Kim \textit{et al.} \cite{kim2024stethoscope} \, (SG-SCL) & \textit{ICASSP`24} & 79.87 & 43.55 & 61.71 & \underline{81.06} & 22.62 & 51.84 \\
    & Kim \textit{et al.} \cite{kim2024bts} \, (Audio-CLAP) & \textit{INTERSPEECH`24} & 80.85 & 44.67 & 62.56 & 70.67 & \textbf{41.90} & \underline{56.29} \\
    \midrule
    \multirow{2}{*}{Multimodal}
    & Kim \textit{et al.} \cite{kim2024bts} \, (BTS) & \textit{INTERSPEECH`24} & \underline{81.40} & \textbf{45.67} & \underline{63.54} & 67.50 & \underline{39.33} & 53.42 \\
    & Ours (BTS-CARD) & - & \textbf{84.42$_{\pm 3.47}$} & \underline{44.83}$_{\pm 2.94}$ & \textbf{64.63$_{\pm 0.57}$} & \textbf{82.02$_{\pm 3.28}$} & \textbf{41.90$_{\pm 4.96}$} & \textbf{61.96$_{\pm 1.50}$} \\
    
    \bottomrule
    \end{tabular}}
    \label{tablex_main_results}
    \vspace{-5mm}
\end{table*}

As shown in Table~\ref{tablex_main_results}, our framework consistently enhances performance in both IND and OOD settings. By contrast, conventional multimodal approaches such as BTS, which fail to account for spurious correlations, perform well under in-distribution but deteriorate markedly under distribution shifts. Unimodal baselines perform even worse, highlighting the efficacy of our debiasing methodology. Overall, the proposed elements contribute to performance improvement across heterogeneous data distributions compared to existing methodologies.

\subsubsection{Ablation Studies}
\label{sec:ablation_studies}

\begin{table}[!t]
\centering
\caption{Ablation studies on IND (ICBHI) and OOD (SPRSound) settings. Components are denoted as: (a) counterfactual debiasing, (b) adversarial debiasing, and (c) counterfactual metadata augmentation. For (b), (c), and Full, the best results across $\alpha$ are reported.}
\scriptsize
\resizebox{\linewidth}{!}{%
\begin{tabular}{ccccccc}
\toprule
\midrule
\multirow{2}{*}{Components} & \multicolumn{3}{c}{IND (In-Distribution)} & \multicolumn{3}{c}{OOD (Out-of-Distribution)} \\
\cmidrule(lr){2-4}\cmidrule(lr){5-7}
 & $S_p$ (\%) & $S_e$ (\%) & Score (\%) & $S_p$ (\%) & $S_e$ (\%) & Score (\%) \\
\hline
\midrule
w/o (a)   & 83.49 & 43.02 & 63.25 & 70.83 & \underline{46.27} & 58.55 \\
w/o (b)   & 82.84 & \textbf{45.94} & \underline{64.33} & \underline{72.50} & \underline{46.27} & \underline{59.39} \\
w/o (c)   & \underline{83.55} & 43.81 & 63.68 & 66.99 & \textbf{47.73} & 56.96 \\
\midrule
Full      & \textbf{84.42} & \underline{44.83} & \textbf{64.63} & \textbf{82.02} & 41.90 & \textbf{61.96} \\
\bottomrule
\end{tabular}
}
\label{tab:ablation_studies}
\vspace{-5mm}
\end{table}

\noindent To assess the contribution of each component, we conduct ablation studies (shown in Table~\ref{tab:ablation_studies}). Removing counterfactual debiasing (w/o (a)) consistently degrades performance, confirming its role in suppressing non-causal associations. Excluding adversarial debiasing (w/o (b)) significantly reduces OOD performance, underscoring its value in mitigating institution-specific biases. Eliminating counterfactual metadata augmentation (w/o (c)) causes the largest OOD drop, highlighting the importance of suppressing metadata-driven shortcuts under distribution shifts.

Across components, our framework slightly reduces $S_e$ while improving $S_p$ in OOD settings. This trade-off arises because debiasing suppresses spurious correlations, such as the over-representation of pediatric cases among healthy samples, making the model less inclined to over-predict healthiness in shifted cohorts. As a result, $S_e$ decreases, but $S_p$ improves, yielding the improved scores.

\subsubsection{Comparative Studies}
\label{sec:comparative_studies}

\begin{table}[!t]
\centering
\caption{Comparative studies of debiasing attributes on IND (ICBHI) and OOD (SPRSound) settings.}
\scriptsize
\resizebox{\linewidth}{!}{%
\begin{tabular}{ccccccc}
\toprule
\midrule
\multirow{2}{*}{Combinations} & \multicolumn{3}{c}{IND (In-Distribution)} & \multicolumn{3}{c}{OOD (Out-of-Distribution)} \\
\cmidrule(lr){2-4}\cmidrule(lr){5-7}
& $S_p$ (\%) & $S_e$ (\%) & Score (\%) & $S_p$ (\%) & $S_e$ (\%) & Score (\%) \\
\hline
\midrule
Age               & \textbf{88.92} & 38.49 & 63.70 & 80.10 & 46.27 & \textbf{63.18} \\
Sex               & 79.61 & \underline{47.24} & 63.42 & 76.44 & 33.16 & 54.80 \\
Location               & 83.72 & 41.38 & 62.55 & \underline{83.65} & 40.10 & 61.88 \\
Device               & 83.22 & 45.11 & \underline{64.17} & 73.37 & \underline{47.04} & 60.20 \\
Age \& Sex        & 77.20 & \textbf{51.06} & 64.13 & 58.56 & \textbf{60.15} & 59.36 \\
Age \& Location \& Device & 81.32 & 45.45 & 63.39 & \textbf{87.02} & 29.31 & 58.16 \\
Location \& Device        & \underline{84.42} & 44.83 & \textbf{64.63} & 82.02 & 41.90 & \underline{61.96} \\
\bottomrule
\end{tabular}
}
\label{tab:comparative_studies}
\vspace{-5mm}
\end{table}

\noindent Next, we investigate the effect of debiasing metadata attributes. Debiasing age leads to moderate improvements for evaluations conducted on both IND and OOD scenarios, with consistently balanced scores. Debiasing sex improves IND stability but considerably degrades OOD performance, suggesting that although this attribute is correlated with label bias, it still conveys information that generalizes across hospitals. Debiasing recording location and device yields relatively balanced outcomes, with strong IND performance while preserving OOD robustness. Further, debiasing age and sex does not generalize well OOD, and debiasing three components of age, location, and device significantly weakens OOD results. Therefore, these findings suggest that our methodology, debiasing both device and location simultaneously, provides the most reliable predictions.

\subsubsection{Parameter Analysis}
\label{sec:parameter_analysis}

\begin{figure}[!ht]
    \centering
    \includegraphics[width=0.4\textwidth]{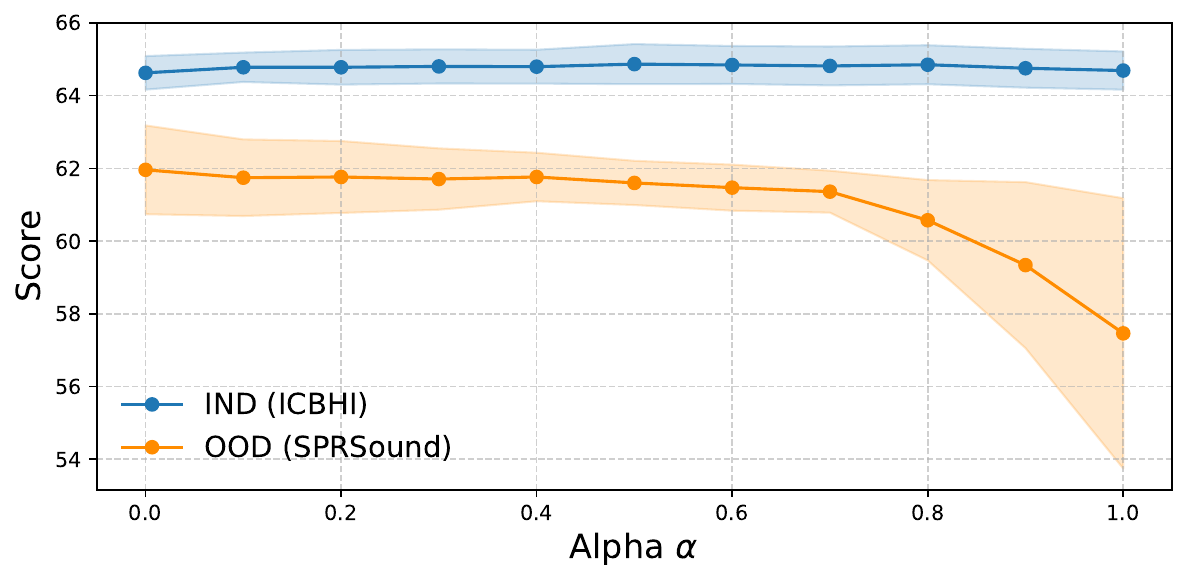}
    \caption{Parameter analysis across different values of $\alpha$ on IND (ICBHI) and OOD (SPRSound) settings.}
    \label{fig:parameter_analysis}
    \vspace{-3mm}
\end{figure}
\noindent We analyze the effect of the debiasing coefficient $\alpha$, which controls the degree to which the NDE is subtracted from the TE during inference. As shown in Figure \ref{fig:parameter_analysis}, IND performance remains relatively stable across $\alpha$, whereas OOD performance deteriorates as $\alpha$ increases. This trade-off indicates that excessive suppression removes informative signals for generalization, prioritizing bias removal over robustness and resulting in saturated IND performance at the cost of degraded OOD generalization.

Interestingly, the strongest generalization is achieved at $\alpha=0$, where inference relies solely on the TE. This suggests that our framework effectively debiases the TE during training, alleviating the need for an explicit trade-off between IND accuracy and OOD robustness at test time. In practice, setting $\alpha=0$ reduces inference costs and eliminates the need to tune an additional hyperparameter, making it a cost-efficient and clinically pragmatic choice.

\section{Conclusion}
\label{sec:conclusion}

In summary, we proposed a novel counterfactual adversarial debiasing framework, BTS-CARD, that effectively mitigates spurious correlations from patient metadata and learns representations that are stable under metadata variation in multimodal RSC. It demonstrated consistent and remarkable improvements over strong baselines under both in-distribution (IND) and out-of-distribution (OOD) settings, thereby enhancing robustness and generalization across different clinical environments.

\newpage

\bibliographystyle{IEEEbib}
\small
\bibliography{refs}

\end{document}